\documentclass[aps,prl,reprint,groupedaddress]{revtex4-1}

\usepackage{graphicx}
\usepackage{epstopdf}
\usepackage{amssymb}
\usepackage{amsmath}
\usepackage{bm}
\usepackage{braket}

\begin{document}

\title{Quantum dynamics with spatiotemporal control of interactions in a stable Bose-Einstein condensate} 

\author{Logan W. Clark, Li-Chung Ha, Chen-Yu Xu, Cheng Chin}
\affiliation{James Franck Institute, Enrico Fermi Institute and Department of Physics, University of Chicago, Chicago, IL 60637, USA}

\date{\today}

\begin{abstract}
Optical control of atomic interactions in a quantum gas is a long-sought goal of cold atom research. Previous experiments have been hindered by short lifetimes and parasitic deformation of the trap potential. Here, we develop and implement a generic scheme for optical control of Feshbach resonance in quantum gases, which yields long condensate lifetimes sufficient to study equilibrium and non-equilibrium physics with negligible parasitic dipole force. We show that fast and local control of interactions leads to intriguing quantum dynamics in new regimes, highlighted by the formation of van der Waals molecules and partial collapse of a Bose condensate. 
\end{abstract}

\maketitle

Spatiotemporal control of interactions would bring a plethora of new quantum-mechanical phenomena into the realm of ultracold atom research. Temporal modulation of interactions is theoretically proposed as a route for creating anyonic statistics in optical lattices \cite{Gong2009, Greschner2014a} as well as new types of quantum liquids \cite{Abdullaev2003, Saito2004} and excitations \cite{Staliunas2002, Ramos2008, Abdullaev2010}. Spatial modulation would grant access to unusual soliton behavior \cite{Rodas-Verde2005, Belmonte-Beitia2007}, controlled interfaces between quantum phases \cite{Hartmann2008}, stable nonlinear Bloch oscillations \cite{Salerno2008}, and even the dynamics of acoustic black holes \cite{Balbinot2008}. The conventional technique for controlling interactions in cold atoms, magnetic Feshbach resonance \cite{Inouye1998, Chin2010}, is typically insufficient for these applications because the magnetic coils are generally too large for very fast or local modulation.

A promising alternative is optical control of Feshbach resonance (OFR). With laser beams, high spatial resolution and high speed control of interactions can be realized by optical modulators. Efforts toward achieving OFR in quantum gases have made significant progress \cite{Fedichev1996, Bohn1997, Fatemi2000, Theis2004, Bauer2009a, Bauer2009b, Zelevinsky2006, Enomoto2008, Yamazaki2010, Blatt2011, Wu2012, Yan2013} but encountered two major obstacles. First, in previous experiments OFR has limited the quantum gas lifetime to the millisecond timescale \cite{Bauer2009b, Yan2013} due to optical excitation to molecular states. Short lifetimes forbid studies of quantum gases in equilibrium or after typical dynamical timescales. Second, the change of interaction strength from OFR is often accompanied by an optical potential. This potential can result in a parasitic dipole force which dominates the dynamics when the interactions are spatially modulated \cite{Yamazaki2010}.

\begin{figure}
 \includegraphics{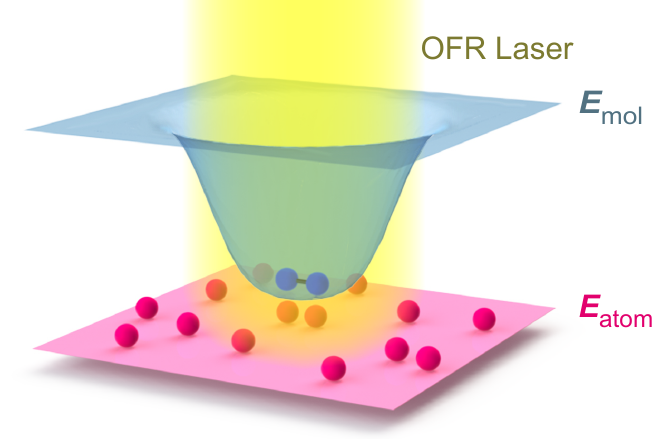}%
 \caption{\label{fig:scheme} \textbf{Illustration of optical control of Feshbach resonances.} A Feshbach resonance occurs when a laser (yellow) brings a molecular energy level (blue surface) close to the atomic scattering threshold (red surface). Here, the atom-molecule coupling makes atomic interactions more attractive at higher laser intensity. Operating at the magic wavelength, the beam does not shift the energy of single atoms (see text).  }
 \end{figure}
 \begin{figure*}
 \includegraphics{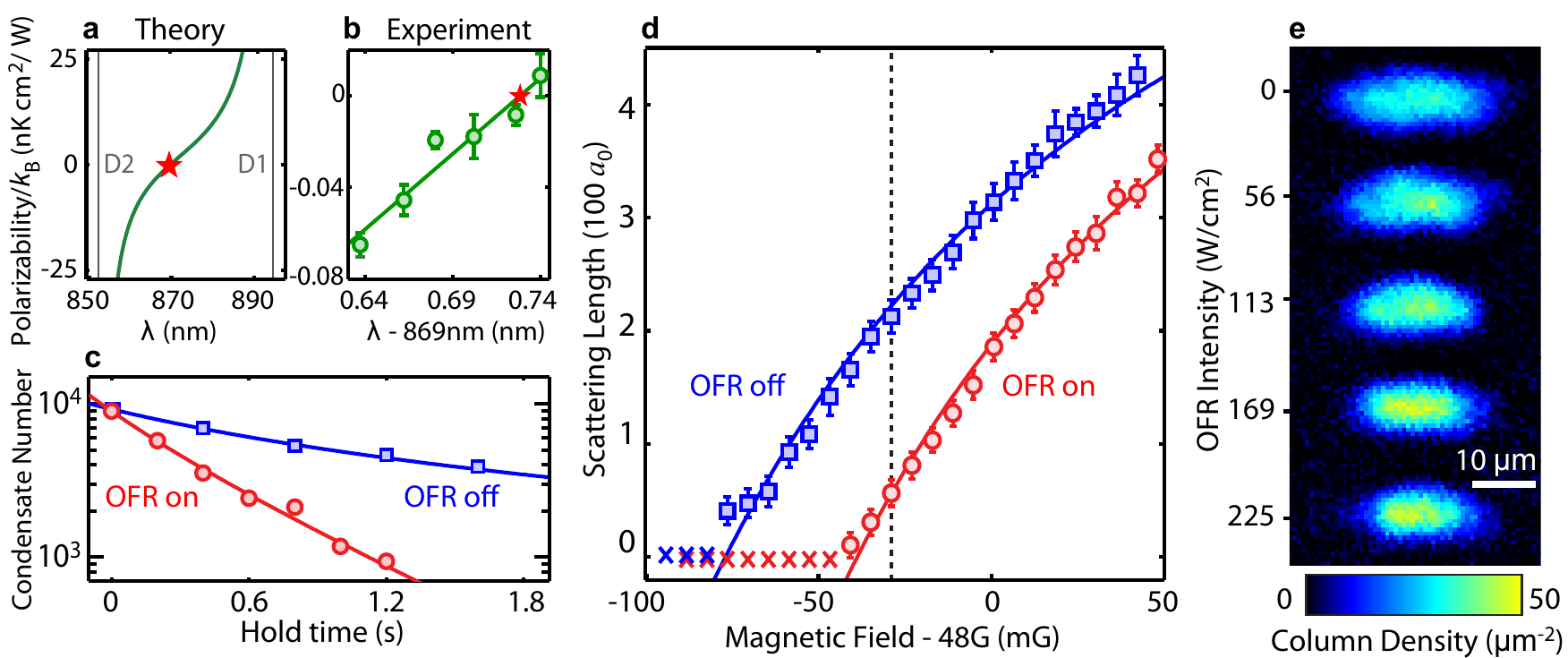}%
 \caption{\label{fig:OFR} \textbf{Stable optical control of scattering length at a magic wavelength $\lambda_\mathrm{M}$.} 
 \textbf{a,} Theoretical polarizability of Cs atoms in the absolute ground state for $\sigma^+$ polarization (supplement). The red $\star$ marks the magic wavelength where polarizability is zero.
 \textbf{b,} Measured polarizability (green $\circ$). A linear fit yields $\lambda_\mathrm{M}=869.73(2)$ nm (red~$\star$).
 \textbf{c,} Number of condensed atoms remaining over time with (red $\circ$) and without (blue $\square$) exposure to OFR at a magnetic field of 48.19~G. We fit the decay dynamics (red and blue curves) and find that OFR exposure adds a one-body loss process with a time constant of 0.63(2)~s (supplement) at intensity $I=225$~W/cm$^2$.
 \textbf{d,} Scattering length $a$ determined from the free expansion of BECs with (red~$\circ$) and without (blue $\square$) exposure to the OFR laser. When the scattering length becomes negative ($a<0$) the condensate collapses ($\times$). The red and blue curves derive from a single fit to all $a>0$ data using Eq.~\ref{eqn:aVsB}, which yields $\Delta=157(3)~$mG, $B_0=47.766(4)~$G, and $\beta I=-38(1)~$mG.
 \textbf{e,} \textit{In situ} images of BECs at $47.97$~G (dashed line in \textbf{d}) after ramping on the OFR intensity over 200~ms. Each image is the average of 10 trials. All error bars show standard error.}
 \end{figure*}

In this report we propose and implement a novel scheme for optically controlling interactions while maintaining long quantum gas lifetime and zero parasitic dipole force. 
With a far detuned laser, a change of the scattering length $a$, which determines the interaction strength, by 180 Bohr radii ($a_0$) is only coupled with a slow radiative loss of (0.63~s)$^{-1}$. This loss rate is two to four orders of magnitude lower than previous reports with a similar change in scattering length and allows the BEC to remain in equilibrium. Furthermore, the laser operates at a magic wavelength to eliminate the atomic dipole potential. We apply this OFR scheme to test the response of BECs to rapid oscillation of interactions down to the timescale of 10~ns, reaching beyond the van der Waals energy scale. By spatially modulating the interactions we observe intriguing dynamics, including the formation and collapse of solitons within a condensate.

We optically control Feshbach resonances by using a far-detuned laser to light shift molecular states near the atomic scattering threshold (FIG. \ref{fig:scheme}). The large detuning from all atomic and molecular transitions offers low heating and loss rates for the quantum gas. For a laser with intensity $I$, the total light shifts of atoms (subscript a) and molecules (subscript m) are given by (see supplement)
\begin{align}
\delta E_\mathrm{a} = &~(\alpha_\mathrm{a} + \beta_\mathrm{a} \mu_\mathrm{a}) I  \nonumber\\
\delta E_\mathrm{m} = &~(\alpha_\mathrm{m} + \beta_\mathrm{m} \mu_\mathrm{m}) I , \nonumber
\end{align}
\noindent  where $\alpha$ is the scalar polarizability and the vector polarizability $\beta \mu$ depends on the magnetic moment $\mu$. Since our target molecular states are very weakly bound, they have similar polarizability to free atoms: $\alpha_\mathrm{m}\approx 2 \alpha_\mathrm{a}$ and $\beta_\mathrm{m} \approx \beta_\mathrm{a} \equiv \beta$ (supplement). Assuming the molecular and atomic magnetic moments differ $\mu_m \neq 2\mu_a$, the vector light shift can bring the molecular states closer to the scattering state, inducing a resonant atom-molecule coupling. Moreover, we choose a magic wavelength $\lambda_\mathrm{M}$ to eliminate the dipole force on the atoms ($\delta E_\mathrm{a}=0$) \cite{Weitenberg2011}, such that only the molecular shift
\begin{equation}
\delta E_\mathrm{m} \approx \beta (\mu_\mathrm{m} - 2 \mu_\mathrm{a}) I
\label{eqn:finalEm}
\end{equation}
remains (FIG. \ref{fig:scheme}). Under these conditions the laser can change the scattering length without creating parasitic dipole forces. This scheme can be implemented in atomic species with a magnetic Feshbach resonance and a magic wavelength far detuned from electronic transitions (supplement).

We implement this OFR scheme in cesium BECs prepared with a small positive scattering length near the Feshbach resonance at 47.8~G (see supplement for experiment details). Theoretically, one possible magic wavelength is $869.7$~nm for a $\sigma^+$ polarized laser, which is far detuned from all electronic transitions but maintains a large vector polarizability (FIG.~\ref{fig:OFR}a). We confirm the magic wavelength experimentally by measuring the dipole force of the OFR laser on the atoms (supplement). From the extracted polarizability, we determine the magic wavelength to be $\lambda_\mathrm{M}=869.73(2)$~nm (FIG.~\ref{fig:OFR}b). At the intensity $I=225~$W/cm$^2$ used for most of this work, we estimate that the residual dipole potential $k_\mathrm{B} \times 1$~nK is negligible compared to our typical chemical potential of $k_\mathrm{B} \times 10$~nK, where $k_\mathrm{B}$ is the Boltzmann constant. With uniform exposure to this intensity, the loss induced by OFR is well explained by a one-body time constant of 0.63(2)~s (FIG.~\ref{fig:OFR}c). 

\begin{figure*}
\includegraphics{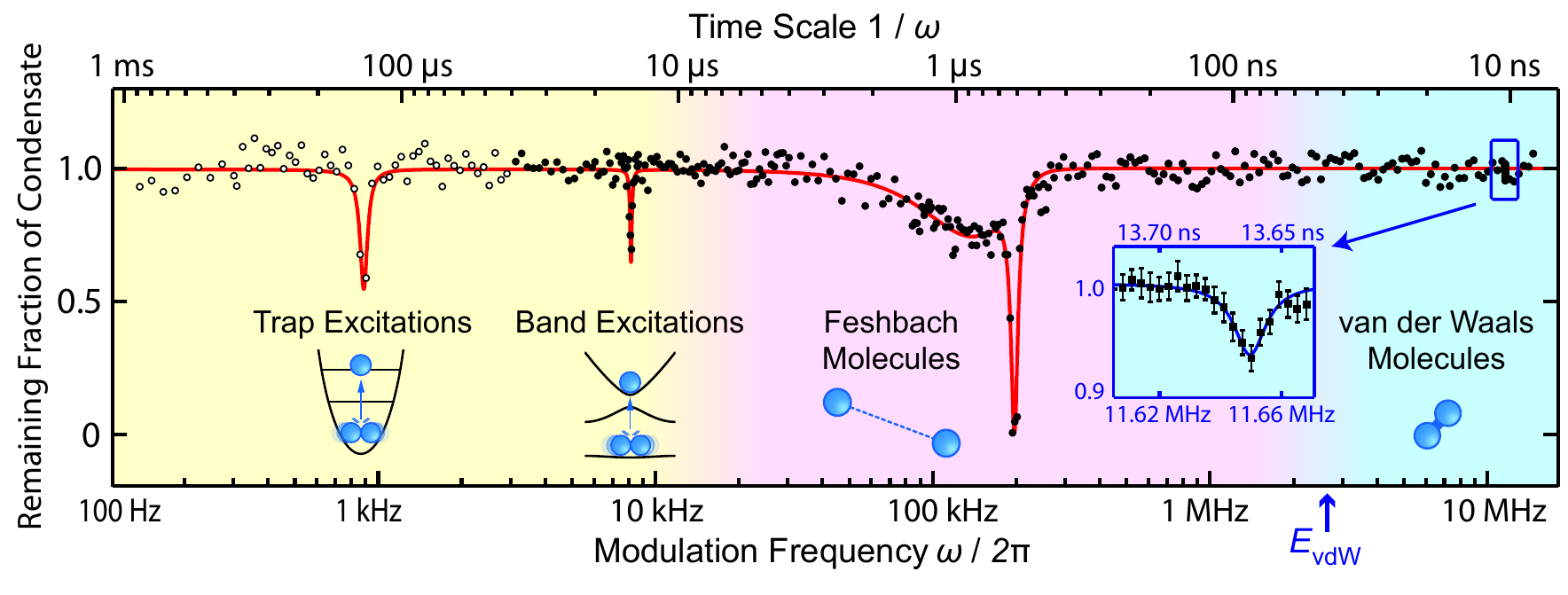} %
\caption{\label{fig:temporal} \textbf{Interaction modulation spectroscopy.} The BEC at $47.976~$G is exposed for a time $t$ to the OFR beam, which is intensity modulated at frequency $\omega/2\pi$. The number of condensed atoms remaining after exposure, normalized to the off-resonant number, is measured under three conditions: ($\circ$) $t=100~$ms with no optical lattice; ($\bullet$) $t=20~$ms with a one dimensional optical lattice of depth $h \times 9.3~$kHz; (inset $\square$) $t=500~$ms with no optical lattice. Resonances are observed at 0.89(1), 8.18(2), 133(7), 197(1), and 11,649(2)~kHz, determined from fits (solid curve) to each resonance (Gaussian for 133~kHz, Lorentzian for others). The illustrations indicate the nature of each resonance. The van der Waals energy scale is $E_{\mathrm{vdW}}=h \times 2.7~$MHz for Cs molecules.}
\end{figure*} 

To precisely determine the change of scattering length, we allow the BEC to freely expand with and without exposure to the OFR laser and measure the size of the gas after expansion \cite{Castin1996, Volz2003} (supplement). FIG.~\ref{fig:OFR}d shows the shift of scattering length induced by the laser near the 47.8~G Feshbach resonance. We fit the scattering lengths with a theoretical model \cite{Chin2010}

\begin{equation}
a(I) = a_\mathrm{bg} \left[1-\frac{\Delta}{B(I) - B_0}\right],
\label{eqn:aVsB}
\end{equation}

\noindent where $B(I) = B_{\mathrm{ex}} + \beta I$ is the effective magnetic field including the OFR contribution $\beta I$ (Eq.~\ref{eqn:finalEm}) and the external field $B_\mathrm{ex}$, $a_{\mathrm{bg}} \approx 950~a_0$ is the background scattering length at this Feshbach resonance \cite{Berninger2013}, $\Delta$ is the width of the resonance, and $B_0$ is the resonance position. The fit yields $\beta I=-38(1)$~mG, sufficient to decrease the scattering length from $180~a_0$ to zero.

The long BEC lifetime allows us to corroborate the change of scattering length based on \textit{in situ} measurements of the density profile. We slowly ramp on the OFR beam to four different final intensities over 200~ms and measure the resulting column density profiles (FIG.~\ref{fig:OFR}e). Higher OFR intensities shrink the BEC and increase its density, consistent with weakening the repulsive interactions. A mean-field analysis yields scattering lengths in excellent agreement with the free expansion measurement (supplement).

The stability of this scheme enables us to explore temporal and spatial control of interactions in a quantum gas. We first perform interaction modulation spectroscopy by recording the response of the BEC to an OFR beam with oscillating intensity, see FIG.~\ref{fig:temporal}. We observe a variety of resonance features over a wide range of timescales, highlighting the versatility of this technique. 

At lower frequencies we observe excitations in the trap and in the optical lattice which directly result from the oscillating scattering length. This demonstrates that plentiful theoretical proposals \cite{Gong2009, Greschner2014a, Abdullaev2003, Saito2004, Abdullaev2010, Staliunas2002, Ramos2008} which require rapid oscillation of scattering length in the bulk or in the lattice are well within reach of our scheme. 

At higher frequencies the oscillating OFR intensity induces formation of molecules. We identify resonances corresponding to a virtual state at 133(7)~kHz, a weakly-bound Feshbach molecular state at 197(1)~kHz, and a deeply-bound van der Waals molecular state at 11.649(2)~MHz. All observed resonance positions are in excellent agreement with theoretical calculations (supplement). These resonances provide direct evidence that OFR can access interaction physics on timescales as short as 10~ns.

\begin{figure}
\includegraphics{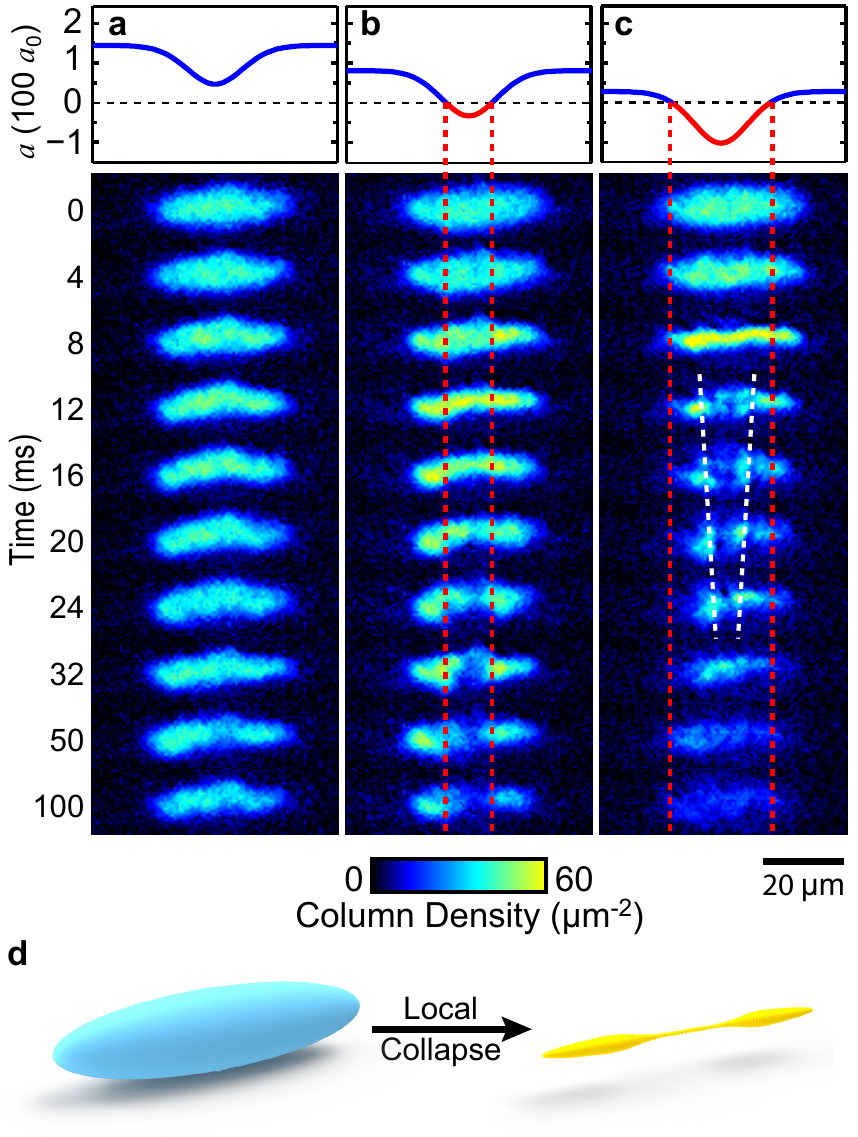} %
\caption{\label{fig:spatial} \textbf{Condensate dynamics with spatially modulated interactions.} Time series of \textit{in situ} images of BECs with $N=12000$ atoms after a quench from uniform $a=200~a_0$ at $47.965$~G to the spatially modulated $a$ shown in the top panels. The OFR beam has peak intensity of $115$~W/cm$^2$ and a waist of $14$~$\mu$m, while the final magnetic fields are \textbf{a},~47.949~G, \textbf{b},~47.935~G, and \textbf{c},~47.925~G. Each image is the average of 6 or 7 trials. The red dashed lines show where $a$ equals zero. The white dashed lines in \textbf{c} guide the eye to the motion of the solitonic wave towards the trap center. \textbf{d,} Illustration of the local collapse dynamics, in which the initial BEC (blue) undergoes transverse compression followed by localized central collapse (yellow).}
\end{figure}

Next, we demonstrate spatial modulation of the interaction strength within a quantum gas. For this experiment we employ an OFR beam which is small compared to the size of the BEC, leading to a reduced scattering length only in the center of the gas. After preparing the BEC we quickly turn on the OFR beam and simultaneously change the magnetic field. We study the subsequent dynamics of the sample by measuring its \textit{in situ} density profile over time (FIG.~\ref{fig:spatial}). For example, when the interactions remain repulsive throughout the condensate (FIG.~\ref{fig:spatial}a), we observe collective excitations for the duration of the experiment. For all images the small distortion at the center of the gas results from the dipole potential due to slightly non-uniform laser polarization (supplement). 

Intriguing quantum dynamics occur when the interactions become locally attractive. When the scattering length is negative in a small region near the center of the trap (FIG.~\ref{fig:spatial}b), we observe a brief period of transverse compression followed by a rapid drop in central density between 20 and 32~ms after the quench, signalling local collapse of the condensate (illustrated in FIG.~\ref{fig:spatial}d). A large fraction of the sample survives at the edges for more than 100~ms. With even stronger attractive interactions (FIG.~\ref{fig:spatial}c), faster central collapse occurs after 8~ms. Subsequently, the dense remnants at the edge of the sample move toward the center of the trap (see white dashed lines in FIG.~\ref{fig:spatial}c). Beyond 32~ms only thermal gas survives, indicating that the remnants have undergone further collapse. Based on this behavior, we identify these remnants as bright matter wave solitons \cite{Cornish2006}, which form at small negative scattering length but become unstable when the scattering length drops below a critical value. The variety of behaviors observed in this experiment establishes the richness of the quantum dynamics accessible with space dependent interactions. 

In conclusion, we implement a generic scheme for optically controlling interactions in quantum gases. This scheme overcomes the key challenges encountered by past approaches to OFR. Fast and local control of interactions in a quantum gas enables studies of novel quantum dynamics and has great potential in the fields of quantum simulation and engineering, opening a new frontier of interaction-driven quantum phenomena.

\subsubsection{Acknowledgements}
We thank Y.T. Chen for assistance in the early stages of the experiment, P.S. Julienne for coupled-channel calculations of Cs molecular structure, and C.V. Parker for helpful discussions. L.W.C. is supported by the NDSEG Fellowship. L.-C.H. is supported by the Grainger Fellowship and the Taiwan Government Scholarship. This work was supported by NSF MRSEC (DMR-0820054), NSF Grant No. PHY-0747907 and ARO-MURI W911NF-14-1-0003.

\bibliography{OFR-arXiv-combined}

\setcounter{equation}{0}
\setcounter{figure}{0}
\setcounter{table}{0}

\renewcommand{\theequation}{S\arabic{equation}}
\renewcommand{\thefigure}{S\arabic{figure}}

\section{Supplementary Information}
\subsection{Experiment Setup}
Our experiments test optical control of Feshbach resonances in almost pure BECs of cesium atoms formed in a crossed beam optical dipole trap (wavelength $\lambda=1064$~nm). This harmonic trap has typical horizontal frequencies $(\omega_x, \omega_y) = 2\pi\times(12, 30)$~Hz and vertical frequency $\omega_z$ controlled by the beam intensities, which we vary from $2\pi\times70$~Hz to $2\pi\times470$~Hz to suit the needs of individual experiments. Our samples typically contain from 3$\sim$10$\times10^3$ atoms with peak densities from 1$\sim$4$\times10^{13}$~cm$^{-3}$. An objective lens (NA=0.5) oriented vertically collects our \textit{in situ} absorption images with $\sim$1~$\mu$m resolution. When we test the response of condensates in optical lattices to oscillating OFR intensity, one of the horizontal trapping beams is retro-reflected to create a 1D optical lattice with spacing $d=532$~nm and depth $h \times 9.28$~kHz. For most of the experiments described in this paper we form our BECs near the $d$-wave magnetic Feshbach resonance at $47.8$~G \cite{Lange2009} with a small and positive scattering length of $a=$~200$\sim$300~$a_0$. The magnetic field in our system is stable to within 1~mG with systematic calibration error of 5~mG, sufficient to form stable BECs.

We optically control interactions with the intensity of the OFR laser, which propagates along the magnetic field direction. The change of scattering length due to OFR is insensitive to small changes in the wavelength and polarization of the OFR laser, however both parameters must be carefully controlled to eliminate the residual light shift. The source of the laser is a free-running single-mode diode with 100~mW output power, which can be tuned via its temperature. We determine the laser frequency using a wavemeter with an accuracy of 100~MHz, sufficient to make the residual light shift negligible.

The temporal modulation experiment (see Fig.~3 in main text) requires rapid control of the beam intensity. We send the OFR beam through an acousto-optic modulator (Isomet 1205C-1) and use a fast RF switch (Mini-Circuits ZFSWA-2-46) to control the acoustic wave which drives the modulator. We measure the 3~dB bandwidth of intensity modulation to be 10~MHz.

\subsection{Theoretical calculation of light shift, scattering rate and performance of OFR}

\begin{figure}
\includegraphics{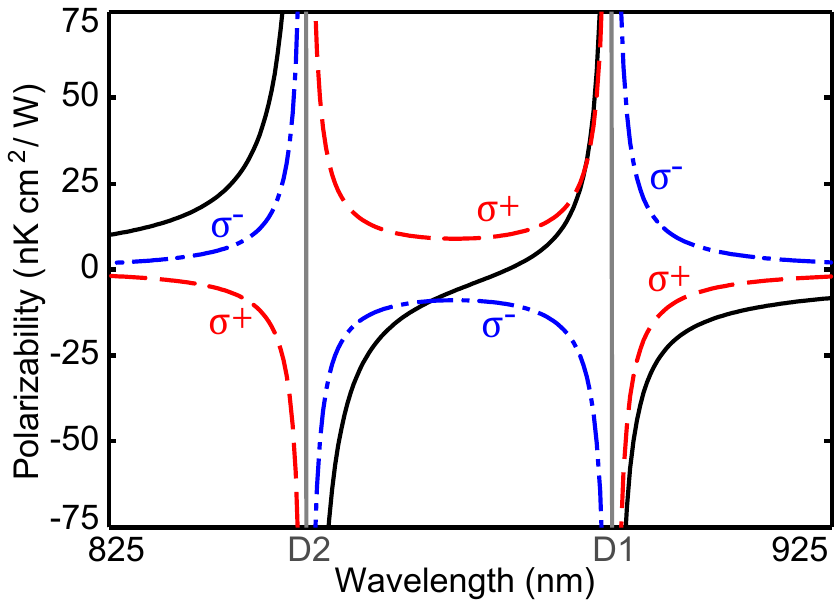} %
\caption{\label{fig:S-Polarizability} \textbf{Theoretical polarizability.} Scalar (black), $\sigma^+$ vector (red dashed), and $\sigma^-$ vector (blue dot-dashed) polarizabilities in the absolute ground state of Cs, see equations \ref{eqn:ScalarPol} and \ref{eqn:VectorPol}. Only the D1 (894~nm) and D2 (852~nm) lines are included in the calculation. Two possible magic wavelengths exist with circularly polarized light, 869.7~nm for $\sigma^+$ and 891.0~nm for $\sigma^-$ polarization. We employ $\sigma^+$ polarization for this work.}
\end{figure}

To predict the magic wavelength $\lambda_\mathrm{M}$ and the effective field shift $\beta I$ from OFR (main text Eq.~1) we calculate the scalar and vector polarizabilities. Our Cs atoms are prepared in the absolute hyperfine ground state $\ket{F=3,~M_F=3}$ where the total angular momentum $F=J+I$ is the sum of electron angular momentum $J=1/2$ and nuclear spin $I=7/2$, and $M_K$ ($K=F,~J$) is the projection onto the quantization axis. At low field we have $\ket{F=3,~M_F=3}=-\sqrt{7/8} \ket{M_J=-1/2,~M_I=7/2}+\sqrt{1/8}\ket{M_J=1/2,~M_I=5/2}$. 

The scalar ($\alpha$) and vector ($\beta\mu$) AC polarizabilities of an atom for detuning which is large compared to the hyperfine splitting are \cite{Kien2012}

\begin{align}
\alpha_i = & \frac{(-1)^{J_i}}{2 \hbar \epsilon_0 c \sqrt{3(2J_i+1)}} \nonumber \\ &\times \sum_{f} (-1)^{J_f} \left\lbrace 
\begin{array}{ccc} 1 & 0 & 1 \\ J_i & J_f & J_i \end{array} \right\rbrace |{\bra{f}}|d|{\ket{i}}|^2 \nonumber \\ & \times\left(\frac{1}{\omega_{fi}-\omega} + \frac{1}{\omega_{fi}+\omega}\right)
\label{eqn:ScalarPol}
\end{align}

\begin{align}
\beta_i \mu_i = &\frac{(-1)^{J_i}}{2 \hbar \epsilon_0 c} \sqrt{\frac{3 J_i}{2 (J_i+1)(2J_i+1)}} \left( \frac{A M_J}{J_i} \right)  \nonumber  \\ &\times\sum_{f} (-1)^{J_f} \left\lbrace 
\begin{array}{ccc} 1 & 1 & 1 \\ J_i & J_f & J_i \end{array} \right\rbrace |{\bra{f}}|d|{\ket{i}}|^2  \nonumber \\ & \times\left(\frac{1}{\omega_{fi}-\omega} - \frac{1}{\omega_{fi}+\omega}\right),
\label{eqn:VectorPol}
\end{align}

\noindent where $\ket{f}$ represents a relevant excited state, $\epsilon_0$ is the vacuum permittivity, $c$ is the speed of light, $\left\lbrace 
\begin{array}{ccc} j_1 &  j_2 & j_3 \\ j_4 & j_5 & j_6 \end{array} \right\rbrace$ is the Wigner 6-j symbol, $|{\bra{f}}|d|{\ket{i}}|$ and $\omega_{fi}$ are the reduced dipole matrix element and resonance frequency for the transition from $\ket{i}$ to $\ket{f}$ (Table~\ref{tab:LineData}), $\omega$ is the laser frequency, $M_J$ is the projection of $J_i$ onto the laser propagation direction, and $A=(I_{\sigma^+} - I_{\sigma^-})/I$ accounts for the beam polarization. We include in the calculation only the D1 and D2 transitions, which dominate in the wavelength range that we consider. The next most significant excited state provides a correction at the $10^{-3}$ level, which is negligible for our purposes. We set $M_J$ to its expectation value of $<{M_J}>=-3/8$ for atoms in the $\ket{F=3,~M_F=3}$ state. The calculated polarizabilities are shown in FIG.~\ref{fig:S-Polarizability}. 

The figure of merit $M$ for choosing the wavelength and polarization of the laser is the ratio of the effective field shift $\beta I$ from OFR to the photon scattering rate $s(I)$,

\begin{align}
M = &\frac{\beta I}{s(I)} \nonumber\\
s(I) = \frac{I}{2 \hbar^2 \epsilon_0 c} &\sum_f \frac{d_{fi}^2 \Gamma_f}{(\omega_{fi} - \omega)^2},
\label{eqn:ScatterRate}
\end{align}

\noindent where $d_{fi}^2 = C_{fi}^\pm |{\bra{f}}|d|{\ket{i}}|^2$ is the squared dipole matrix element, $C_{fi}^\pm$ is a numerical factor primarily accounting for the Clebsch-Gordan coefficients with beam polarization $\sigma^{\pm}$, and $\Gamma_f$ is the spontaneous emission rate of the excited state $\ket{f}$ (Table~\ref{tab:LineData}). A larger absolute value of $M$ indicates a greater shift in molecular states for a fixed quantum gas lifetime.

$M$ is optimized by using pure circular polarization. The magic wavelengths for $\sigma^+$ and $\sigma^-$ polarization are both suitable choices with almost identical $M$ to within $5\%$. They remain differentiated by the direction of the change in scattering length, $da/dI<0$ for $\sigma^+$ and $da/dI>0$ for $\sigma^-$. We employ $\sigma^+$ polarization with $M=-130$~mG$\cdot$s at the magic wavelength $\lambda_\mathrm{M}=869.7$~nm.

\begin{table*}
\begin{tabular}{| r | c |c |c| c | c|} 
 \hline
& ~~~$\omega_{fi}/2\pi$~(THz)~~~&~~~$|{\bra{f}}|d|{\ket{i}}|~(10^{-29}~\mathrm{C}\cdot$m)~~~&~~~$\Gamma_f/2\pi$~(MHz)~~~&~~~$C_{fi}^+$~~~&~~~$C_{fi}^-$~~~\\
\hline
  D1~ & 335.116 & 3.81 & 4.557 & $\frac{7}{24}$ & $\frac{1}{24}$ \\
  \hline
 D2~ & 351.726 & 5.36 & 5.219 & $\frac{5}{48}$ & $\frac{11}{48} $ \\
 \hline
\end{tabular}
\caption{\label{tab:LineData} The parameters used to calculate the polarizability and scattering rate for the D1 and D2 transitions of Cs.}
\end{table*}

Optimal performance of our OFR scheme is obtained when a laser with maximum $M$ is employed near a proper magnetic Feshbach resonance. The sensitivity of scattering length to the OFR shift (based on Eq. 2 of the main text) is
\begin{equation}
\frac{da}{d(\beta I)}\bigg|_{|a/a_{\mathrm{bg}}|\ll1} = \frac{a_{\mathrm{bg}}}{\Delta}
\end{equation}
\noindent near the zero crossing $B(I) = B_0 + \Delta$ where the condensate can be stable. This result suggests that a narrow resonance with a large background scattering length will offer wide tunability for a given laser intensity. However, one must be cautious to avoid extremely narrow resonances which can greatly enhance the three-body recombination rate. In this work we choose the Feshbach resonance at $47.8$~G, for which we have $a_{\mathrm{bg}}/\Delta=6.6~a_0$/mG from coupled channel calculations \cite{Berninger2013}. 

Overall, optimal performance is obtained via the independent optimizations of $M$, which is determined by the laser, and $da/d(\beta I)$, which is determined by the choice of Feshbach resonance. In our system the product of these two factors yields $M (a_\mathrm{bg}/\Delta)=$$-$860~$a_0\cdot$s. This value indicates that a decrease in scattering length by 860~$a_0$ should be possible with a scattering rate of 1/s.

\subsection{Implementing OFR with other atomic species}

Our scheme for OFR is quite general and should perform well with a variety of atomic species. For example, calculations based on $^{87}$Rb in the absolute ground state $\ket{F=1, M_F=1}$ yield the figure of merit $M=-45$~mG$\cdot$s. Moreover, past experiments with $^{87}$Rb suggest two potential candidates for the Feshbach resonance. For $^{87}$Rb in $\ket{F=1, M_F=1}$, the $1007.4$~G resonance \cite{Bauer2009a} has a sensitivity of $da/d(\beta I)=0.5~a_0$/mG \cite{Chin2010}, which would allow a change of scattering length by 100~$a_0$ with a $5/$s one-body scattering rate. If a magic wavelength is not required, then the interstate Feshbach resonance at 9.13~G \cite{Widera2004} offers $da/d(\beta I)=6.7~a_0$/mG, where $a$ is the scattering length between the $\ket{F=1,~M_F=1}$ and $\ket{F=2,~M_F=-1}$. This resonance allows a change of 300~$a_0$ with a $1~$s$^{-1}$ scattering rate, enabling many interesting studies with optical control of interstate scattering. 

Beyond alkali atoms, quantum gases of heavy, magnetic atomic species like erbium and dysprosium \cite{Lu2011, Aikawa2012} are promising candidates for OFR. Their rich sets of optical transitions provide many options for obtaining a favorable ratio of the vector light shift to the scattering rate. Moreover, such highly magnetic elements have an abundance of Feshbach resonances \cite{Frisch2014, Baumann2014}, many of which could provide favorable properties for implementing OFR.

\subsection{Measurement of polarizability}

We determine the polarizability near the magic wavelength by measuring the displacement of the BEC caused by the OFR laser. For this experiment we operate at $B=22$~G where the dependence of interactions on the OFR shift is negligible. We prepare the OFR beam with a waist of 70~$\mu$m, which is large compared to the BEC, and displace the beam to achieve a slope of $dI/dx=5$~(W/cm$^2$)/$\mu$m across the sample. A finite total polarizability $\gamma = \alpha+\beta \mu$ leads to a force of $F=\gamma dI/dx$ on the atoms, which offsets the center of the harmonic trap by $\Delta x = F / m \omega_x^2$. By measuring the shift in the center of the BEC, we extract the polarizability

\begin{equation}
\gamma = \frac{m \omega_x^2 \Delta x}{dI/dx},
\end{equation}

\noindent yielding the data shown in FIG.~2b of the main text. 

We find the magic wavelength to be $\lambda_M=869.73(2)$~nm for $\sigma^+$ polarization, in fair agreement with the prediction of $869.66$~nm. We attribute the difference to the electronic transitions not included in the calculation and the imperfect beam polarization; we estimate that $I_{\sigma^-}\approx0.005I$. 

In addition, at the magic wavelength we test the assumption that $\beta_m\approx\beta_a\equiv\beta$ for weakly-bound molecular states. We measure the effective field shift $\beta I$ induced by the laser using microwave spectroscopy. We compare the microwave result to the shifts in energy of $g$-wave and $d$-wave molecular states. Within 5\% all methods yield values which are consistent with the theoretical prediction (Eq.~\ref{eqn:VectorPol}).

\subsection{Extracting scattering length from BECs after free expansion}

A precise way to measure the scattering length is to release the BEC from the harmonic trap, simultaneously switch the magnetic field and OFR beam intensity, and measure the BEC radius after a period of free expansion. This method yields measured radii $R$ which scale as $a^{1/2}$, making it more sensitive than \textit{in situ} measurements or free expansion measurements with constant magnetic field which yield $R \propto a^{1/5}$ \cite{Volz2003}. Ref.~\cite{Castin1996} shows that, during expansion, the BEC density profile is parabolic with time-dependent Thomas-Fermi radii $R_k$ ($k=x, y, z$). The time evolution of the ratios $\lambda_k(t) \equiv R_k(t)/R_k(0)$ is described by the three coupled differential equations,

\begin{equation}
\ddot{\lambda_k} =  \frac{a_f}{a_i} \frac{\omega_k^2}{\lambda_x \lambda_y \lambda_z \lambda_k}
\label{eqn:S-FreeExpansion}
\end{equation}

\noindent where $\omega_k$ is the trap frequency before release, $a_i=250a_0$ is the scattering length before release, and $a_f$ is the scattering length during expansion which is determined by the magnetic field and OFR intensity. Note that there is one equation for each $\lambda_k$ and the trap frequencies are independently calibrated, such that the only unknown variable is the scattering length.

\begin{figure}
\includegraphics{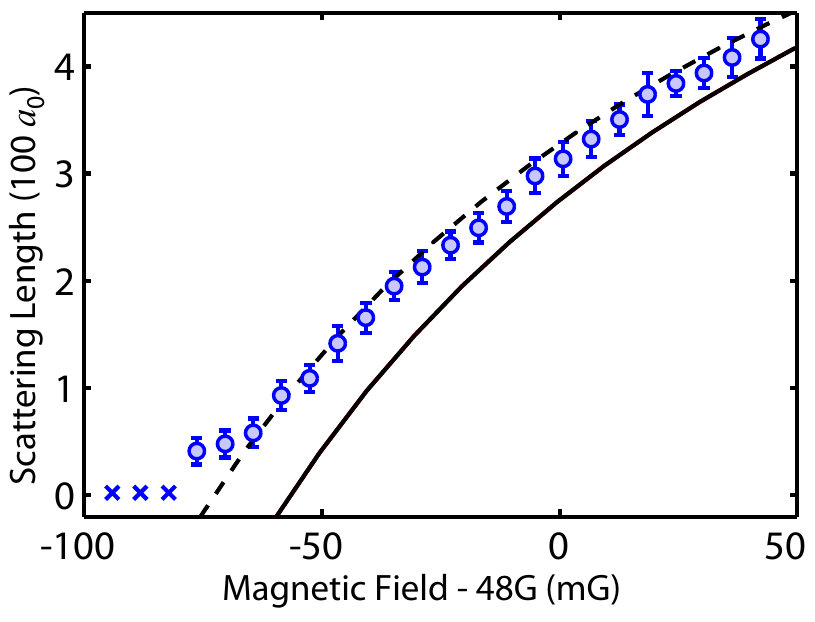} %
\caption{\label{fig:S-ScatteringLength} \textbf{Comparison of scattering length from free expansion to coupled channel model.} The scattering length extracted from free expansion with no OFR laser (blue $\circ$) compared to the scattering length based on the coupled channel calculations in Ref. \cite{Berninger2013} (solid curve). The dashed curve shows the calculated scattering length shifted by -16~mG, which is within the model uncertainty.}
\end{figure}

To determine $a_f$ as a function of magnetic field and OFR intensity, we execute the experimental procedure over a range of magnetic fields with and without exposure to the OFR beam at $I=225$~W/cm$^2$. For this experiment we use a vertical trapping frequency of $\omega_z=2\pi\times105$~Hz much greater than the horizontal frequencies of $(\omega_x, \omega_y)=2\pi\times(14, 31)$~Hz, which causes the vertical expansion to dominate after release. Thus, after an expansion time of 16~ms we measure only the vertical radius of the gas. For each combination of magnetic field and OFR intensity, we determine the scattering length $a_f$ by comparing the measured radius to the numerical solution of Eq.~\ref{eqn:S-FreeExpansion}. This procedure yields the results shown in FIG.~2d of the main text. In the absence of the OFR beam we compare our results to coupled channel calculations \cite{Berninger2013}, see FIG.~\ref{fig:S-ScatteringLength}.

\subsection{Extraction of scattering length from \textit{in-situ} BEC density profiles}

\begin{figure}
\includegraphics{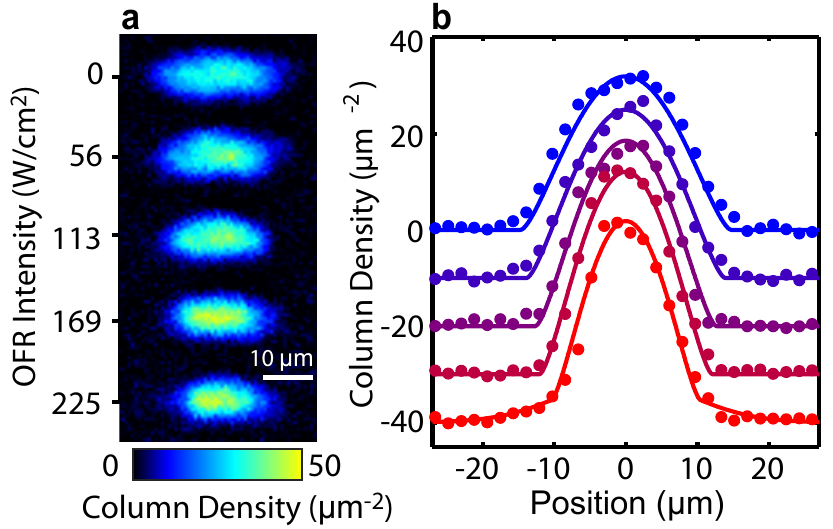} %
\caption{\label{fig:S-InSitu} \textbf{Effect of OFR on in situ density profiles.} \textbf{a,} The in situ image series from fig.~2e of the main text. \textbf{b,} Horizontal line cuts ($\bullet$) obtained by averaging the middle five pixels from each image in \textbf{a}, similarly arranged from top (no OFR laser) to bottom (225~W/cm$^2$). The solid lines are the corresponding cuts from a fit to each image with Eq.~\ref{eqn:S-Density2D}. For the 225~W/cm$^2$ case we first perform a Gaussian fit to the wing to account for thermal atoms before fitting the remaining BEC. The fitted scattering lengths, in the order of increasing beam intensity, are $a$ = 230(33), 175(25), 120(16), 96(13), and $78(16)~a_0$ in good agreement with the free expansion result.}
\end{figure}

We fit the measured \textit{in-situ} column density of BECs with a Thomas-Fermi density profile \cite{pethick2002}

\begin{equation}
n(x,y) = n_0\left(1-\frac{x^2}{R_x^2}-\frac{y^2}{R_y^2}\right)^{3/2},
\label{eqn:S-Density2D}
\end{equation}

\noindent where $n_0=\left(\frac{1}{2\pi}\right)(5 N \omega_x \omega_y)^{3/5} \left(\frac{m^2}{3 \hbar^2 a \omega_z}\right)^{2/5}$ is the peak column density, and $R_k =  \omega_k^{-1}(15 \hbar^2 m^{-2} N \omega_x \omega_y \omega_z a)^{1/5}$. The only free parameter is the scattering length $a$ which appears in $n_0$ and $R_k$, while the atom number $N=3500$ and trap frequencies $(\omega_x, \omega_y, \omega_z) = 2\pi\times(10, 22, 75)$~Hz are calibrated independently. We compare horizontal line cuts of the images to the corresponding cuts from the fit in FIG.~\ref{fig:S-InSitu}.

\subsection{Condensate lifetime with OFR}
In the presence of the OFR laser, we expect the lifetime to be limited by one-body off-resonant scattering of photons, since our scheme does not rely on proximity to resonant atomic or molecular transitions. 
In practice, while we predict the one-body loss process to have extremely weak wavelength dependence, we find significant wavelength dependence of the lifetime within the 40~GHz tuning range of our laser. We attribute this observation to the two-body loss caused by photoassociation or molecular resonances which happen to be within the tested wavelength range. After finding a wavelength which minimizes this loss, we make minute adjustments to our laser polarization to shift the magic wavelength to the loss minimum. The required change in polarization is small, such that its effect on the effective field shift is negligible.

Having minimized two-body loss, we model the lifetime by accounting for the one-body loss process induced by OFR and the pre-existing three-body loss process. With both processes the decay of density $n$ is described by the differential equation

\begin{equation}
\frac{\partial n(\bm{r},t)}{\partial t} = -L_1 n(\bm{r},t) - L_3 n(\bm{r},t)^3,
\label{eqn:densityDecay}
\end{equation}

\noindent where $L_1$ ($L_3$) is the one-(three-)body decay constant. We assume that the loss is slow such that the system remains in equilibrium. Integrating the Thomas-Fermi density profile \cite{pethick2002} over space with fixed scattering length and absorbing trap-dependent factors into the modified decay constant $L_3'$ we obtain,

\begin{equation}
\frac{dN}{dt} = -L_1 N - L_3' N^{9/5}
\label{eqn:totalDecay}
\end{equation}

\noindent which describes the decay of total atom number in the presence of one- and three-body loss. Note that the one-body lifetime is independent of trap geometry.

To determine the lifetime experimentally, we compare the decay of condensate number with and without exposure to the OFR laser at $B_{\mathrm{ex}}=48.19$~G (main text FIG.~2c). We chose this field to obtain realistic decay rates near the Feshbach resonance but be far enough away to avoid inducing condensate dynamics or significantly altering the three-body loss with the OFR beam. We keep the trap depth constant throughout the decay, so that the BEC continues to evaporate and maintain an approximately constant temperature while exposed to the laser. With no OFR exposure we fit the decay to the numerical solution of Eq.~\ref{eqn:totalDecay} with negligible $L_1$ and find $L_3'=5.6 \times 10^{-4}$~s$^{-1}$ in our trap geometry. We use this fixed value to fit the decay in the presence of OFR to Eq.~\ref{eqn:totalDecay} and obtain $L_1=1.58(4)$~s$^{-1}$, corresponding to a time constant of $\tau=1/L_1=0.63(2)~$s, with OFR intensity $225$~W/cm$^2$. 

\begin{figure}
\includegraphics{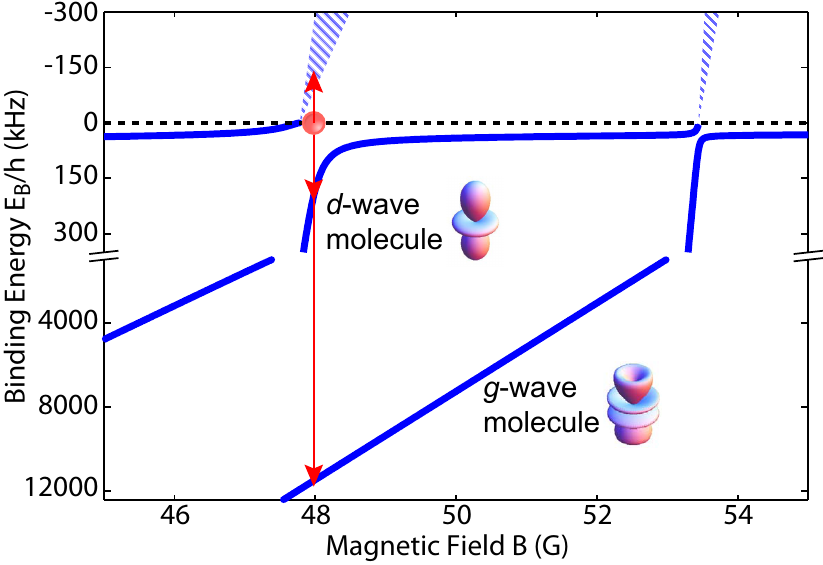} %
\caption{\label{fig:S-ChengBindingEnergy} \textbf{Energy levels for Cs$_2$ molecules.} A schematic of the three molecular resonances observed in our experiment, based on the energy levels in ref.~\cite{Lange2009}. The red ball indicates the condensate, and arrows represent the coupling to three molecular states by interaction modulation spectroscopy.}
\end{figure}

The fitted time constant represents two orders of magnitude improvement over existing OFR schemes, but it is still shorter than the off-resonant scattering time constant of 3.2~s (Eq.~\ref{eqn:ScatterRate}). We expect that the lifetime is shorter than the scattering time constant because the recoil heating from a single scattering event is $k_B\times200$~nK, much larger than the critical temperature $T_C\approx30$~nK of the condensate. Thus each scattering event can lead to the loss of multiple atoms from the condensate. There may also be a contribution from photoassociation resonances or molecular transitions which remain near the chosen wavelength.

\subsection{Theory comparisons for temporal modulation resonances}

We have determined the origin of each of the resonances excited by interaction modulation spectroscopy (main~text~FIG.~3).
The first resonance at 890(10)~Hz corresponds to excitation in the vertical harmonic trap. Since the BEC density is even across the trap center, the oscillating interaction strength provides an even perturbation which can only excite the gas to states with the same parity as the ground state; thus the first excited state is forbidden. The ratio of the resonance frequency to the vertical trapping frequency of 470~Hz is 1.9, suggesting that the samples are close to the quasi-2D regime (chemical potential $\mu\ll$~470~Hz) and OFR is driving excitations to the 2nd excited harmonic oscillator state.

The second resonance at 8.18(2)~kHz corresponds to excitations to the second excited band of the optical lattice. For comparison we perform band structure calculations for our lattice depth of $h\times9.28$~kHz, yielding energies of $h\times6.58$~kHz and $h\times8.18$~kHz for the zero quasimomentum states in the first and second excited bands, respectively. The observed resonance is consistent with transitions to the second excited band; no resonance appears at the first excited band energy because the parity of the ground state is again conserved. Transitions to higher parity-allowed bands are also not observed, which we attribute to the weaker coupling strength to those states.

The highest three resonances correspond to coupling of the free atoms to Cs$_2$ molecular states. The process is analogous to the binding of dimers with an oscillating magnetic field \cite{Thompson2005}. The locations of the observed features are consistent with known Cs$_2$ molecular states (FIG.~\ref{fig:S-ChengBindingEnergy}). We compare our data to coupled channel calculations \cite{Berninger2013} at the average effective magnetic field $B = 47.957$~G. The 133(7)~kHz resonance corresponds to a molecular state above the atomic threshold, with an energy of 95~kHz based on extrapolation of coupled channel calculations into the continuum. The feature is broad because the molecular state is embedded in the continuum. The 197(1)~kHz resonance corresponds to a primarily $d$-wave bound state calculated to be at 208~kHz and the 11.649(2)~MHz resonance corresponds to a $g$-wave bound state calculated at 11.731 MHz \cite{Berninger2013}. These molecules are individually stable, but their collisions with each other and the remaining atoms lead to net loss of atoms from the trap.

\end{document}